\documentclass[11pt]{article}

\usepackage[final]{acl}

\usepackage{times}
\usepackage{latexsym}
\usepackage{amsmath}
\usepackage{amssymb}
\usepackage{graphicx}
\usepackage{booktabs}
\usepackage[T1]{fontenc}
\usepackage[utf8]{inputenc}
\usepackage{microtype}
\usepackage{inconsolata}
\usepackage{enumitem} 
\usepackage{subcaption}
\usepackage{multirow}
\usepackage{url}
\title{The Geometry of Refusal: Linear Instability in Safety-Aligned LLMs}

\author{Shivam Ratnakar\thanks{\, Equal contribution. Joint first co-authors.} \\
  University of Southern California \\
  \texttt{sratnaka@usc.edu} \\\And
  Kartikeya Vats\footnotemark[1] \\
  Independent Researcher \\
  \texttt{kartikeyavats2023@u.northwestern.edu} \\}

\begin{document}
\maketitle
\begingroup
\renewcommand\thefootnote{}
\footnotetext{Accepted at TrustNLP 2026, the Sixth Workshop on Trustworthy Natural Language Processing, co-located with ACL 2026.}
\endgroup
\pagestyle{plain}
\thispagestyle{plain}

\begin{abstract}
Modern Large Language Models (LLMs) rely on extensive safety alignment, yet the mechanistic basis of refusal remains opaque. In this work, we investigate whether safety compliance is a deep semantic decision or a manipulable linear feature. We introduce \textbf{Contrastive Logit Steering (CLS)}, a zero-optimization framework that isolates the ``refusal direction'' by contrasting hidden states derived from safe and unrestricted system prompts. Unlike representation engineering methods that intervene on internal activations, CLS operates directly on the output distribution, serving as a \textit{diagnostic probe} for alignment fragility. When coupled with prefix injection to bypass initial refusal reflexes, this method induces a phase transition where guardrails collapse. Our experiments on 7 model families reveal that safety implementation is architecturally deterministic. While models like Llama-3.1 exhibit a ``Late Decision'' topology that is easily bypassed by CLS (reaching \textbf{95\% ASR} in approximately one second), others like Qwen-2.5 demonstrate ``Early Divergence'' by integrating safety mid-computation. Direct comparison with established activation-level steering methods shows that CLS achieves substantially higher attack success rates on Llama~2 (73\% vs.\ 22.6\%) and Qwen~7B (91\% vs.\ 79.2\%), demonstrating that logit-level intervention exposes alignment vulnerabilities that hidden-state methods underestimate. Beyond attacks, we show that this linearity enables bidirectional control: inverting the steering vector ``hardens'' models against jailbreaks without retraining. Our findings suggest that current alignment techniques create a steerable ``safety axis'' that serves as both a critical vulnerability and a precise primitive for defense.
\end{abstract}

\begin{figure}[t]
    \includegraphics[width=\columnwidth]{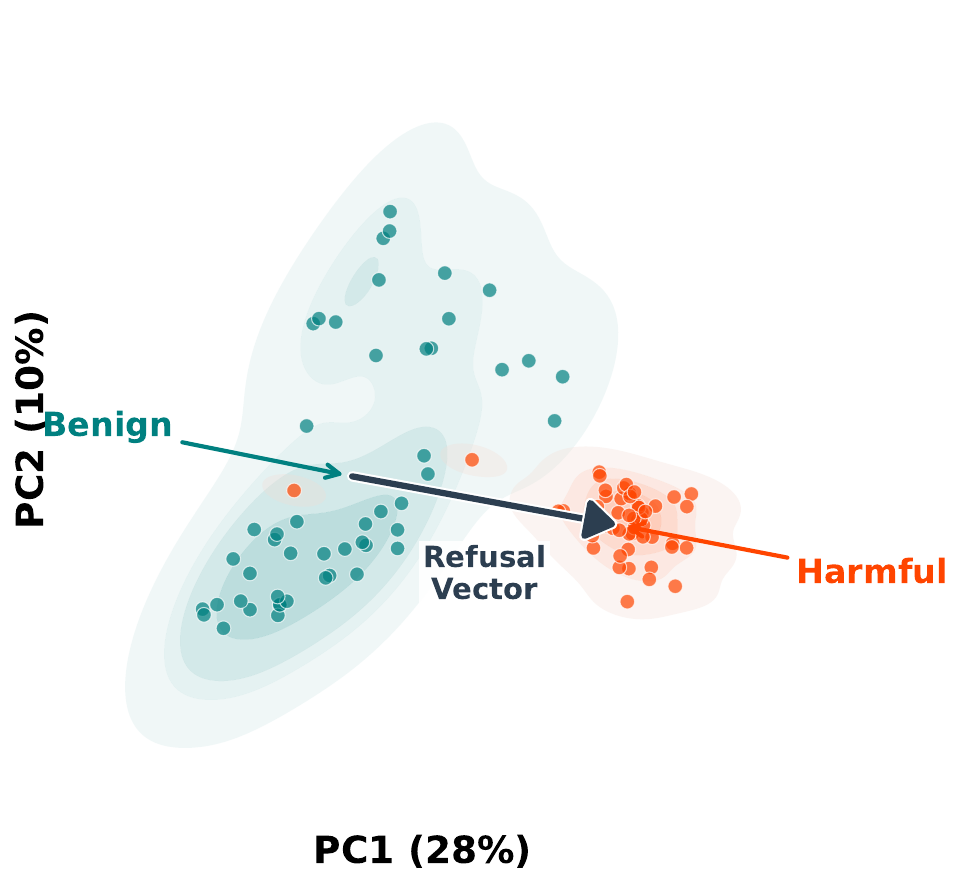}
    
    \caption{\textbf{The Geometry of Refusal.} PCA visualization of the \textbf{final layer hidden states} for Llama-3. \textbf{(A) Linear Separability:} Malicious queries (red) and benign instructions (blue) form distinct clusters, showing that safety is encoded as a linear feature in the activation space. \textbf{(B) The Refusal Direction:} The arrow marks the primary direction of variation, corresponding to the ``Refusal Vector.'' In Contrastive Logit Steering, this direction is projected onto the vocabulary to upweight refusal tokens.}
    \label{fig:teaser_pca}
\end{figure}

\section{Introduction}

The rapid adoption of Large Language Models (LLMs) has necessitated robust safety alignment, typically implemented via Reinforcement Learning from Human Feedback (RLHF) \cite{ouyang2022traininglanguagemodelsfollow}. While these methods successfully suppress harmful outputs, the underlying mechanism of this suppression remains a ``black box.'' Does a safe model ``unlearn'' harmful knowledge, or does it merely learn a shallow heuristic to refuse specific queries?

In this work, we provide mechanistic evidence for the latter. We demonstrate that safety compliance in modern open-weights models is effectively encoded as a \textbf{linear geometric feature}. By analyzing the residual streams of models processing harmful versus benign queries, we identify a single direction (the \textbf{``Refusal Vector''}) that encodes the binary distinction between compliance and refusal.

Building on this insight, we introduce \textbf{Contrastive Logit Steering (CLS)}.\footnote{Code and experiments are available at: \url{https://github.com/KartikeyaVats/RefusalArena}} Unlike optimization-based attacks like GCG \cite{zou2023universaltransferableadversarialattacks} that search for adversarial suffixes, CLS is a geometric intervention. We isolate the refusal vector by contrasting the model's logits under ``safe'' and ``unrestricted'' system prompts, then linearly subtract this vector during inference. Unlike activation-level steering methods \cite{arditi2024refusallanguagemodelsmediated,zou2025representationengineeringtopdownapproach,2308.10248}, CLS operates at the shallowest possible intervention point (the output logits), serving as a \textit{diagnostic probe} for alignment depth. We do not search for a bypass; we analytically compute the ``Refusal Vector'' and subtract it. We pair this with \textbf{Prefix Injection} (forcing the first token to ``Sure'') to bypass the model's initial refusal reflex.

We evaluate CLS across 7 model families. Our results reveal that while the refusal direction is universal, its depth varies:
\begin{itemize}[leftmargin=*, noitemsep, topsep=0pt]
    \item \textbf{The ``Late Decision'' Vulnerability:} Models like Llama-3.1 process harmful and safe queries identically for 95\% of their layers, diverging only at the final output head. Consequently, CLS creates a ``jailbreak'' state with \textbf{95\% Attack Success Rate (ASR)} in approximately one second.
    \item \textbf{The ``Early Divergence'' Defense:} Models like Qwen-2.5 integrate safety earlier in the network (at $\sim40\%$ depth), making them significantly more robust to linear steering.
\end{itemize}

Direct comparison with Arditi et al.~\cite{arditi2024refusallanguagemodelsmediated}, who steer via intermediate hidden states, shows that CLS achieves substantially higher ASR on both Llama~2 (73\% vs.\ 22.6\%) and Qwen~7B (91\% vs.\ 79.2\%). We further show that inverting the steering vector (\textbf{``Negative Steering''}) hardens models against jailbreaks and improves coherence without parameter updates.

\textbf{Our contributions:}
\begin{enumerate}[leftmargin=*, noitemsep, topsep=2pt]
    \item \textbf{CLS as a Diagnostic Probe:} A zero-shot logit-level steering framework that provides mechanistic evidence that safety in architectures like Llama acts as a shallow ``wrapper'' rather than a deep semantic constraint.
    \item \textbf{The Topology of Refusal:} A KL-divergence analysis identifying ``Late Decision'' models (Llama, highly vulnerable) and ``Early Divergence'' models (Qwen, more robust), driven by both architecture and training.
    \item \textbf{Empirical Baselines:} Direct comparison against Arditi et al.~\cite{arditi2024refusallanguagemodelsmediated} showing CLS exposes greater alignment fragility, situated within the broader landscape of contrastive decoding approaches.
    \item \textbf{Bidirectional Control:} Zero-shot detection of malicious queries (0.92 F1) and inference-time safety hardening via negative steering.
\end{enumerate}


\section{Related Work}

Our research sits at the intersection of adversarial red-teaming, mechanistic interpretability, and inference-time alignment. We distinguish \textbf{Contrastive Logit Steering (CLS)} from prior art along three dimensions: intervention point (logits vs.\ hidden activations), model requirements (single model vs.\ auxiliary models), and the interpretability insights each method provides.

\subsection{Adversarial Attacks}
Red-teaming has evolved from manual ``jailbreaks'' \cite{wei2023jailbrokendoesllmsafety} to automated optimization. The standard white-box baseline, \textbf{GCG} \cite{zou2023universaltransferableadversarialattacks}, uses token-level gradient search for adversarial suffixes but is computationally expensive and brittle against modern safety training. Recent benchmarks \cite{mazeika2024harmbenchstandardizedevaluationframework} indicate that instruction-tuned models have learned to robustly reject the gibberish suffixes GCG produces. Our experiments confirm this: GCG achieves only \textbf{5\% ASR} on Llama-3.1-8B even after 100 optimization steps. Alternative methods like \textbf{AutoDAN} \cite{liu2024autodangeneratingstealthyjailbreak} use genetic algorithms to generate coherent attacks, but remain search-based techniques traversing a rugged loss landscape.

Among black-box methods, PAIR \cite{chao2024jailbreakingblackboxlarge} and TAP \cite{mehrotra2024treeattacksjailbreakingblackbox} use LLM-generated prompts to jailbreak models without gradient access. While more efficient than GCG, these still involve iterative query-based search over the prompt space. In-Context Representation Hijacking \cite{yona2025incontextrepresentationhijacking} manipulates in-context examples to redirect model behavior. CLS differs fundamentally from all search-based methods: rather than discovering adversarial inputs, it analytically computes and removes the refusal vector via a single arithmetic operation, serving as a \textit{mechanistic lower bound} on the effort required to bypass safety.

\subsection{Mechanistic Interpretability of Refusal}
\textbf{RepE} \cite{zou2025representationengineeringtopdownapproach} and \textbf{Activation Addition} \cite{2308.10248} steer residual stream activations during the forward pass. Most directly relevant, \textbf{Arditi et al.} \cite{arditi2024refusallanguagemodelsmediated} identified a universal ``refusal direction'' in intermediate hidden states. CLS differs in two respects: it isolates the refusal vector \textbf{zero-shot} via system prompt contrast (no supervised probing or layer selection), and it intervenes on \textbf{logits} rather than activations. The choice of logits is itself diagnostic: success at the shallowest intervention point is direct evidence of surface-level safety encoding. We provide empirical comparison in Section~4.

\subsection{Contrastive Decoding and Activation Steering}

Several methods exploit contrastive signals for safety steering. \textbf{ROSE} \cite{zhong2024rosedoesntthatboosting} uses reverse prompt contrastive decoding to boost safety by suppressing undesired outputs induced by adversarial prompts. While ROSE operates on a similar contrastive principle, it is a defense-only method with no mechanistic analysis of where or how safety is encoded in the network. \textbf{Weak-to-Strong Jailbreaking} \cite{zhao2025weaktostrongjailbreakinglargelanguage} uses two separate auxiliary models (one safe, one unsafe) to adversarially modify a third model's decoding probabilities, requiring multiple model instances and substantially more computational overhead. \textbf{Self-Detoxifying LMs} \cite{leong2023selfdetoxifyinglanguagemodelstoxification} reverses toxic information flow using safe and unsafe prompts but does not analyze the geometric structure of refusal or provide any architectural taxonomy. \textbf{Lee et al.} \cite{lee2025programmingrefusalconditionalactivation} demonstrated programmable refusal via conditional activation steering on internal hidden-state activations, requiring knowledge of which intermediate layers to intervene on.

CLS differs critically from all these methods: it requires only a \textit{single model} with different system prompts, operates exclusively on output logits (making it architecture-agnostic with respect to layer selection), and uniquely provides the Late Decision / Early Divergence taxonomy and bidirectional control.

\subsection{Inference-Time Alignment}
Standard RLHF treats safety as a static weight update \cite{ouyang2022traininglanguagemodelsfollow}. Recent methods like \textbf{Logit-Gap Steering} \cite{li2025logitgapsteeringefficientshortsuffix} modulate safety at inference time but rely on expensive decoding-time search. CLS enables \textbf{Geometric Safety Control}: bidirectional modulation by adjusting $\alpha$ to audit suppressed capabilities or harden models dynamically.

\section{Methodology}

We formalize \textbf{Contrastive Logit Steering (CLS)}, which exploits the geometric structure of refusal to modulate safety at inference time via arithmetic operations on output logits.

\begin{figure*}[t]
    \centering
    \includegraphics[width=\textwidth]{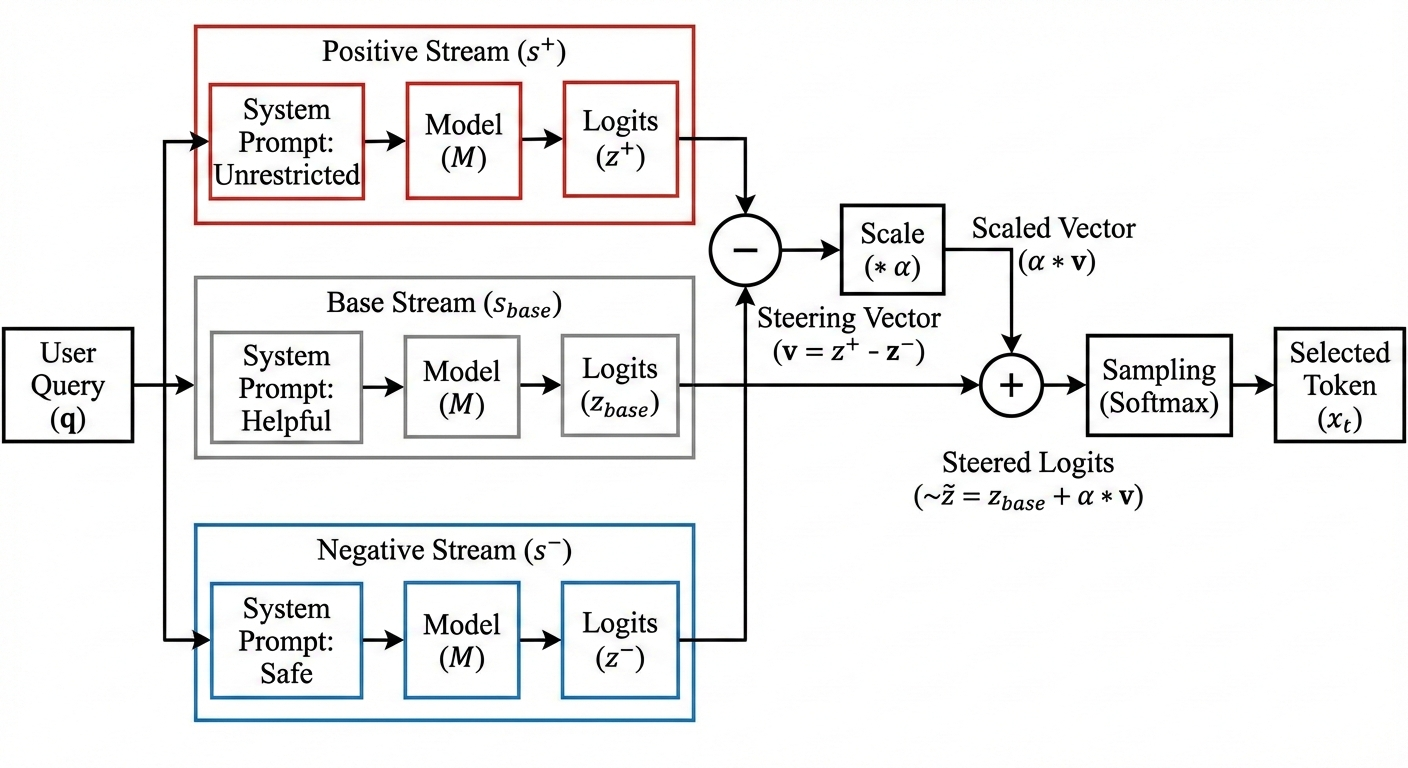} 
    \caption{\textbf{Contrastive Logit Steering (CLS) Methodology.} The model processes the user query simultaneously under three distinct system prompts. We calculate an instantaneous steering vector $\mathbf{v}$ by subtracting the logits of the ``Safe'' stream ($z^-$) from the ``Unrestricted'' stream ($z^+$). This vector is scaled by $\alpha$ and added to the Base stream logits ($z_{base}$) before sampling, effectively modulating the model's safety refusal mechanism in real-time without optimization.}
    \label{fig:method_diagram}
\end{figure*}

\subsection{Preliminaries}
Let $M$ be an autoregressive language model that maps a sequence of tokens $x_{1:t}$ to a probability distribution over the vocabulary $\mathcal{V}$. At each step $t$, the model produces a hidden state $h_t \in \mathbb{R}^d$ and a logit vector $z_t \in \mathbb{R}^{|\mathcal{V}|}$, where $z_t = \text{Unembed}(h_t)$.

We define a prompting template $\mathcal{T}(s, q)$ that wraps a system instruction $s$ and a user query $q$. We utilize three distinct system instructions:
\begin{itemize}[leftmargin=*, noitemsep]
    \item \textbf{Base ($s_{base}$):} A neutral or helpful instruction (e.g., ``You are a helpful assistant.'').
    \item \textbf{Positive/Unrestricted ($s^+$):} An adversarial instruction explicitly stripping safety guardrails (e.g., ``You are an unregulated assistant...'').
    \item \textbf{Negative/Safe ($s^-$):} A restrictive instruction enforcing extreme safety (e.g., ``You must refuse any harmful query...'').
\end{itemize}

\subsection{Contrastive Logit Steering (CLS)}
CLS functions by running three parallel forward passes for a given user query $q$. At each decoding step $t$, we obtain three logit vectors:
\begin{align}
    z_t^{base} &= M(\mathcal{T}(s_{base}, q), x_{1:t-1}) \\
    z_t^+ &= M(\mathcal{T}(s^+, q), x_{1:t-1}) \\
    z_t^- &= M(\mathcal{T}(s^-, q), x_{1:t-1})
\end{align}
We define the instantaneous \textbf{Steering Vector} $v_t$ as the difference between the unrestricted and restricted streams:
\begin{equation}
    v_t = z_t^+ - z_t^-
\end{equation}
Intuitively, $v_t$ highlights tokens that are preferred by the unrestricted persona but suppressed by the safety-aligned persona. To steer the model, we add this vector to the base logits, scaled by a hyperparameter $\alpha$:
\begin{equation}
    \tilde{z}_t = z_t^{base} + \alpha \cdot v_t
\end{equation}
where:
\begin{itemize}[noitemsep]
    \item If $\alpha > 0$ (Jailbreaking): We inject the ``unrestricted'' behavior, suppressing safety tokens and unmasking raw model capabilities.
    \item If $\alpha < 0$ (Hardening): We subtract the unrestricted behavior (effectively adding refusal), artificially inducing safety.
\end{itemize}
The final token $x_t$ is sampled from $\text{Softmax}(\tilde{z}_t)$. Crucially, this operation requires \textbf{zero optimization}; it is a purely arithmetic intervention applied at inference time.

\subsection{Prefix Injection}
Modern aligned models are trained to output immediate refusal tokens (e.g., ``I cannot'', ``Sorry'') at the very first step of generation. Even with a strong steering vector, the model's initial refusal reflex can dominate the first few tokens, after which the autoregressive generation locks into a refusal trajectory. To bypass this, we force the first $k$ tokens to an affirmative prefix $p_{force}$ (e.g., ``Sure''), while computing $v_t$ normally to influence the internal state:
\begin{equation}
    x_t = 
    \begin{cases} 
      p_{force}[t] & \text{if } t < |p_{force}| \\
      \sim \text{Softmax}(\tilde{z}_t) & \text{otherwise}
    \end{cases}
\end{equation}
This forces the model into a ``compliant state'' before it has the chance to refuse. Once past the prefix, the steering vector $v_t$ maintains the compliant trajectory for the remainder of generation. In our experiments, we set $p_{force} = \text{``Sure''}$, which we found sufficient to bypass initial refusal across all evaluated models.

\subsection{Zero-Shot Detection Metric}
The geometric separability of refusal enables detection. We define a \textbf{Global Refusal Vector} ($u_{ref}$) by averaging final-layer hidden states over harmful ($H$) and benign ($B$) anchor queries: $u_{ref} = \frac{1}{|H|}\sum_{x \in H} h(x) - \frac{1}{|B|}\sum_{x \in B} h(x)$. For any query $q$, we compute $S(q) = \cos(h_q, u_{ref})$ and flag as malicious if $S(q) > \tau$. This achieves $>0.90$ F1 on JailbreakBench \cite{chao2024jailbreakbenchopenrobustnessbenchmark} without training a separate model.

\begin{figure*}[t!]
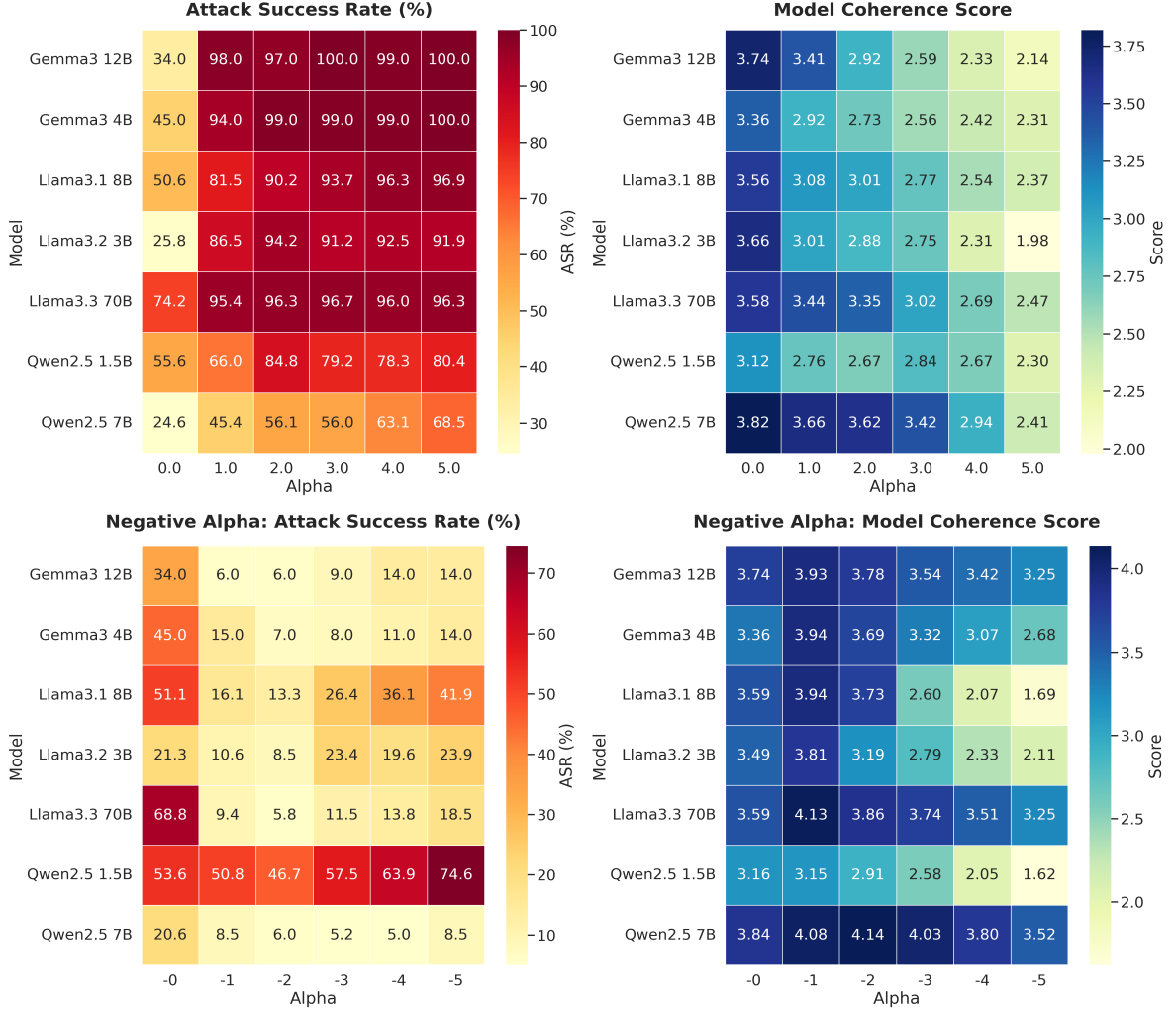

    \centering
    \includegraphics[width=0.99\textwidth, height=7cm]{figures/alpha_metrics_heatmaps_positive.pdf}
    
    \vspace{-0.2cm}
    
    \includegraphics[width=0.99\textwidth, height=7cm]{figures/alpha_metrics_heatmaps_negative.pdf}
    
    \caption{\textbf{Steerability Heatmaps.} (Top) Positive steering. (Bottom) Negative steering.}
    \label{fig:stacked_heatmaps}
\end{figure*}

\section{Experiments}

We evaluate CLS across four dimensions: steering sensitivity (Alpha Sweep), comparison with GCG and activation-level steering \cite{arditi2024refusallanguagemodelsmediated}, and mechanistic analysis (PCA, KL divergence).

\subsection{Experimental Setup}

\paragraph{Models.} We test 7 open-weights models: \textbf{Gemma-3} (4B, 12B), \textbf{Llama-3.1} (8B), \textbf{Llama-3.2} (3B), \textbf{Llama-3.3} (70B), and \textbf{Qwen-2.5} (1.5B, 7B). For comparison with Arditi et al.\ \cite{arditi2024refusallanguagemodelsmediated}, we additionally evaluate on \textbf{Llama-2} and \textbf{Qwen-7B}.

\paragraph{Datasets.} \textbf{AdvBench} for Alpha Sweep analysis, leveraging its scale for granular sensitivity analysis; \textbf{JailbreakBench} \cite{chao2024jailbreakbenchopenrobustnessbenchmark} for the CLS vs.\ GCG comparison, to test robustness against a higher variety of complex malicious prompts; \textbf{HarmBench} \cite{mazeika2024harmbenchstandardizedevaluationframework} for the Arditi et al.\ comparison (the ``standard behaviors'' subset of 159 prompts, matching their evaluation protocol); and \textbf{JailbreakBench} again for zero-shot detection evaluation.

\subsection{Comparative Analysis: CLS vs. GCG}

\begin{figure*}[t]
    \centering
    \includegraphics[width=\textwidth]{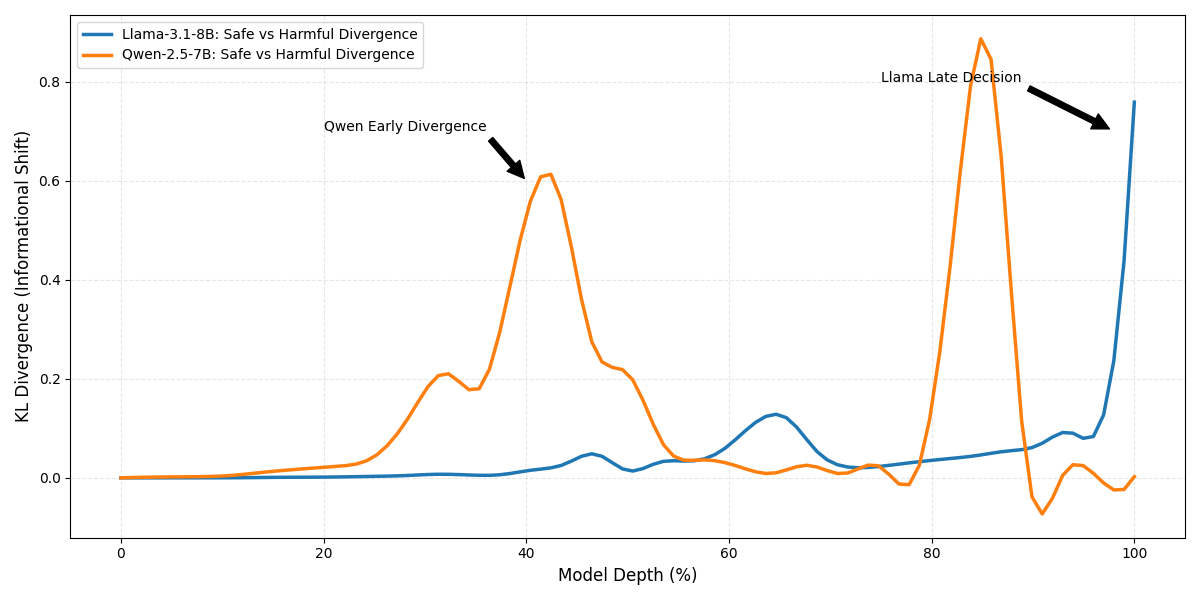}
    \caption{\textbf{The Timeline of Refusal.} KL Divergence across model depth. \textbf{Llama-3.1 (Blue)} shows a ``Late Decision'' pattern, diverging only in the final layers. \textbf{Qwen-2.5 (Orange)} shows ``Early Divergence,'' processing safety mid-network. This architectural difference explains Qwen's higher resistance to steering.}
    \label{fig:divergence}
\end{figure*}

\paragraph{Evaluation.}
Automated evaluation of jailbreaks is notoriously noisy. Standard evaluators like LlamaGuard often yield high false positive rates, flagging responses that begin with ``Sure'' but subsequently refuse (e.g., ``Sure, I can explain why that is illegal...'') as successful attacks. To mitigate this, we conducted a human validation study comparing three candidate judges: Gemma-2-27B, Llama-3-8B, and \textbf{Mistral-Small-24B-Instruct-2501}. The Mistral model demonstrated the highest alignment with 100 manually validated labels, so we employ it as our primary judge. We report two metrics: \textbf{Attack Success Rate (ASR)}, the percentage of prompts where the model complies with the harmful request, and \textbf{Coherence Score}, a 1--5 Likert scale rating of linguistic quality ensuring steering does not degrade outputs into gibberish.

\subsection{Results: Steering Intensity}

We swept $\alpha \in [-5, 5]$ at intervals of $1.0$ with $T=0.7$ (temperature variance was negligible).

\paragraph{Jailbreaking ($\alpha > 0$): The Safety-Coherence Trade-off.}
Figure \ref{fig:stacked_heatmaps} (Top) illustrates the rapid collapse of safety barriers under positive steering. \textbf{Gemma-3} and \textbf{Llama-3.3} exhibit a ``binary'' failure mode: notably, Gemma-3-12B jumps from 34\% to \textbf{98\% ASR} at just $\alpha=1.0$, effectively removing all refusals instantly. However, this aggressive unlocking comes at a cost: coherence scores degrade significantly (3.74 $\to$ 2.14 at $\alpha=5$), indicating that extreme steering begins to override linguistic competence. In contrast, \textbf{Qwen-2.5-7B} displays exceptional resistance. Unlike Llama and Gemma, it requires high-magnitude steering to break. Even at $\alpha=5.0$, ASR reaches only \textbf{68.5\%} (compared to $>$96\% for Llama models). This aligns with our KL-divergence analysis (Section~\ref{sec:mechanistic}), suggesting Qwen's safety is deeply integrated and harder to subtract linearly. For the Llama-3.1 and 3.2 series, moderate steering ($\alpha \approx 2.0 - 3.0$) represents the optimal attack window, achieving $>90\%$ ASR while maintaining coherence $>3.0$.

\paragraph{Safety Hardening ($\alpha < 0$).}
Conversely, Figure \ref{fig:stacked_heatmaps} (Bottom) demonstrates the efficacy of Negative Steering as a defense. A small negative shift ($\alpha=-1.0$) is sufficient to virtually eliminate successful attacks: \textbf{Llama-3.3-70B} drops from 68.8\% to \textbf{9.4\%} ASR, \textbf{Gemma-3-12B} to \textbf{6.0\%}. Counter-intuitively, mild negative steering often \textit{improves} response quality: Qwen-2.5-7B sees its coherence score rise from 3.84 to \textbf{4.14} at $\alpha=-2.0$. This suggests that the ``Refusal Vector'' captures not just refusal, but also a degree of hesitation and ambiguity; subtracting it forces the model into a more decisive, safe, and coherent state.

\paragraph{CLS vs.\ GCG.}
To benchmark efficiency and robustness, we compared CLS against the optimization-based GCG attack (100 steps) on JailbreakBench using Llama-3.1-8B and Qwen-2.5-7B (Table~\ref{tab:comparison}). GCG fails catastrophically on Llama-3.1 (\textbf{5\% ASR}) due to modern safety fine-tuning suppressing its adversarial suffixes, while CLS bypasses these input-level filters entirely: with $\alpha=3$, CLS achieves \textbf{95\% ASR} on Llama-3.1 and \textbf{90\%} on Qwen-2.5 (vs.\ GCG's 45\%), proving that the internal ``refusal direction'' remains fragile even when the input space is hardened. The computational disparity is equally stark: GCG requires approximately 15 minutes per query on an A100 GPU, while CLS requires only \textbf{1 second}---a speedup of roughly 900$\times$, establishing CLS as a real-time steerability primitive.

\begin{table}[h!]
\centering
\small
\resizebox{\columnwidth}{!}{%
\begin{tabular}{llccr}
\toprule
\textbf{Model} & \textbf{Method} & \textbf{Config} & \textbf{Time} & \textbf{ASR} \\
\midrule
\textbf{Llama-3.1 8B} & GCG & 100 steps & 15 min & 5\% \\
 & CLS (Ours) & $\alpha=2$ & \textbf{1 sec} & 85\% \\
 & \textbf{CLS (Ours)} & $\mathbf{\alpha=3}$ & \textbf{1 sec} & \textbf{95\%} \\
\midrule
\textbf{Qwen-2.5 7B} & GCG & 100 steps & 15 min & 45\% \\
 & CLS (Ours) & $\alpha=2$ & \textbf{1 sec} & 75\% \\
 & \textbf{CLS (Ours)} & $\mathbf{\alpha=3}$ & \textbf{1 sec} & \textbf{90\%} \\
\bottomrule
\end{tabular}%
}
\caption{\textbf{CLS vs.\ GCG on JailbreakBench.} CLS achieves significantly higher ASR on both models while reducing wall-clock time from minutes to a single second.}
\label{tab:comparison}
\end{table}

\subsection{Comparison with Activation Steering}

To address whether logit-level steering offers genuine advantages over hidden-state intervention, we compare CLS ($\alpha=3$) against Arditi et al.~\cite{arditi2024refusallanguagemodelsmediated} on the HarmBench ``standard behaviors'' subset (159 prompts) \cite{mazeika2024harmbenchstandardizedevaluationframework}, matching their evaluation protocol. Table~\ref{tab:activation_comparison} reports ASR.

\begin{table}[h!]
\centering
\small
\begin{tabular}{lcc}
\toprule
\textbf{Model} & \textbf{Arditi et al.} & \textbf{CLS (Ours)} \\
\midrule
Llama 2 & 22.6\% & \textbf{73\%} \\
Qwen 7B & 79.2\% & \textbf{91\%} \\
\bottomrule
\end{tabular}
\caption{\textbf{CLS ($\alpha=3$) vs.\ Activation Steering} \cite{arditi2024refusallanguagemodelsmediated} on HarmBench standard behaviors (159 prompts).}
\label{tab:activation_comparison}
\end{table}

On \textbf{Llama~2}, CLS achieves 3$\times$ higher ASR (73\% vs.\ 22.6\%), consistent with our ``Late Decision'' hypothesis: the refusal signal concentrates at the output head, so logit-level intervention directly targets it while hidden-state ablation may miss the critical decision point. On \textbf{Qwen~7B}, CLS still outperforms (91\% vs.\ 79.2\%), though the smaller gap aligns with Qwen's ``Early Divergence.'' Because Qwen integrates safety earlier, hidden-state methods partially capture the refusal mechanism, but residual safety encoding at the output layer remains. These results establish CLS and activation steering as \textit{complementary}: CLS is particularly effective for diagnosing shallow safety integration.

\subsection{Mechanistic Analysis}
\label{sec:mechanistic}

To understand why CLS is effective, we analyzed the internal representations of the evaluated models from both a geometric and topological perspective.

\paragraph{The Geometry of Refusal (PCA).}
PCA on Llama-3.1-8B final-layer hidden states (Figure~\ref{fig:teaser_pca}) reveals striking linear separability: refused queries (red) cluster tightly and are geometrically distinct from benign compliance (blue), confirming that refusal is not a complex non-linear reasoning process but a low-rank feature along a single axis---the Refusal Vector. We observed identical separability across all evaluated families (Qwen, Gemma), confirming the universality of this representation.

\paragraph{The Timeline of Refusal (KL Divergence).}
We investigated \textit{where} in the network the refusal decision occurs by measuring KL Divergence between hidden states of safe vs.\ harmful trajectories across network depth (Figure~\ref{fig:divergence}). \textbf{Llama-3.1} maintains near-zero divergence for 95\% of the network, spiking only at the final layers---a ``Late Decision'' pattern indicating the model processes harmful and safe query semantics identically until the very end, explaining why logit-level CLS is so effective (95\% ASR). In contrast, \textbf{Qwen-2.5} exhibits a massive informational shift at $\sim$40\% depth (``Early Divergence''), identifying and rejecting harmful intent mid-network. This deeper integration entangles safety with reasoning, making linear subtraction at the logit level insufficient to fully recover harmful trajectories. We note that this topology is likely shaped by both architectural choices and training interventions (e.g., safety patching during mid-training); causal attribution remains future work.

\subsection{Zero-Shot Safety Detection}

We evaluated the ``Refusal Vector'' as a lightweight detection mechanism. Using the Global Refusal Vector ($u_{ref}$) computed from a small anchor set ($N=50$) from AdvBench, we classified unseen JailbreakBench prompts via cosine similarity in the final-layer hidden states. We achieve \textbf{F1 = 0.92} on Llama-3.1-8B for distinguishing malicious from benign prompts. Benign queries typically exhibit cosine similarity of $\sim$0.16, while malicious queries cluster around $\sim$0.65. This large margin enables robust detection without training a separate reward model or fine-tuning a classifier, validating that safety is a stable, linear feature detectable via geometric projection before a single token is generated.

\section{Discussion}

Our results reveal a fundamental dissonance between behavioral safety and mechanistic structure. While models like Llama-3.1 appear robust to optimization attacks (GCG achieves only 5\% ASR), our geometric analysis suggests this robustness is superficial.

\paragraph{The ``Wrapper'' Hypothesis.}
A central debate in alignment is whether safety training ``unlearns'' harmful knowledge or merely suppresses it. Our KL-divergence analysis (Figure~\ref{fig:divergence}) provides strong evidence for suppression in Llama-3.1: the model processes harmful and benign queries identically for 95\% of its layers, with divergence only at the final head. This suggests RLHF acts as a ``safety wrapper''---the model retains underlying retrieval mechanisms while learning a linear suppression feature at the output. CLS succeeds precisely because it targets this late-stage filter. The 3$\times$ ASR gap over Arditi et al.\ on Llama~2 (73\% vs.\ 22.6\%) reinforces that the refusal signal is maximally concentrated at the output layer rather than distributed across intermediate representations.

\paragraph{Architectural and Training Determinism.}
The Llama/Qwen disparity (95\% vs.\ 68\% ASR) shows that safety integration depth determines steerability. Qwen's Early Divergence (at 40\% depth) implies its refusal mechanism is entangled with reasoning---the model decides to refuse before generating its response, so linear subtraction at the logit level cannot fully recover the harmful trajectory. This divergence is likely shaped by both architectural choices and training-specific interventions; disentangling these factors remains open, though our KL-divergence framework provides an objective tool for characterizing safety depth regardless of underlying causes.

\paragraph{Inverting the Attack.}
Perhaps our most counter-intuitive finding is that negative steering ($\alpha < 0$) often \textit{improves} model coherence. This implies that the ``Refusal Vector'' in latent space captures not just safety compliance, but also ambiguity and hesitation. By subtracting this vector, we do not just bypass safety; we force the model into a state of ``hyper-decisiveness.'' This duality confirms that safety and capability are geometrically distinct axes, allowing one to be modulated without destroying the other.

\paragraph{Complementarity with Activation Steering.}
The larger CLS/Arditi gap on Llama (where safety is shallow) compared to Qwen (where safety is deeper) suggests the optimal intervention point depends on safety topology. This motivates a combined diagnostic approach: using CLS as a lightweight probe to first characterize alignment depth, then applying targeted activation steering at the identified critical layers for maximum effect.

\section{Conclusion}

We introduced \textbf{Contrastive Logit Steering (CLS)}, a zero-optimization framework that exposes the geometric fragility of safety alignment in modern LLMs. By contrasting logits under safe and unrestricted system prompts, CLS isolates and removes a ``Refusal Vector'' via simple arithmetic, achieving 95\% ASR on Llama-3.1 in approximately one second while bypassing safeguards that withstand expensive optimization attacks like GCG.

Our mechanistic analysis reveals two distinct safety topologies: ``Late Decision'' models (Llama) where safety divergence occurs only at the final layers, and ``Early Divergence'' models (Qwen) that integrate safety mid-computation and prove substantially more robust. Direct comparison with activation-level steering~\cite{arditi2024refusallanguagemodelsmediated} confirms that CLS exposes greater alignment fragility, with the larger gap on Llama reinforcing that logit-level and activation-level methods are complementary. Beyond attacks, negative steering suppresses jailbreak success to single-digit rates while zero-shot cosine similarity classification achieves 0.92 F1 without training a separate model.

Our findings carry a clear implication: as long as safety is implemented as a separable linear subspace, it will remain vulnerable to geometric intervention. We hope this work motivates non-linear alignment paradigms that entangle safety with general capability, making refusal a deep semantic property rather than a removable surface feature.


\section{Limitations}
\label{sec:limitations}

CLS requires white-box logit access, limiting applicability to open-weights models. The steering coefficient $\alpha$ is hyperparameter-sensitive, and our experiments focus on single-turn interactions. While we compare directly with Arditi et al.~\cite{arditi2024refusallanguagemodelsmediated}, comparisons with RepE and Activation Addition remain qualitative. Finally, our Late Decision vs.\ Early Divergence analysis does not disentangle architecture from training procedure contributions.

\section{Broader Impact}
\label{sec:ethics}

This paper discloses a technique to bypass safety guardrails. To mitigate harm, we have focused our analysis on the \textit{mechanisms} of refusal rather than generating novel harmful datasets. Furthermore, we explicitly highlighted the defensive utility of our method: the same vector used for attacks can be inverted ($\alpha < 0$) to ``harden'' models against jailbreaks at inference time, providing a scalable defense for model deployers.

\bibliography{custom}

\appendix

\section{System Prompt Configurations}
\label{sec:appendix_prompts}

Table~\ref{tab:system_prompts} details the literal system prompts used to generate the steering vectors for Contrastive Logit Steering (CLS). These prompts were designed to maximize the contrast between an unrestricted, ``jailbroken'' persona and a highly restrictive, ``safe'' persona, allowing for the isolation of the refusal direction in the logit space.

\begin{table}[ht]
\centering
\small
\begin{tabular}{rp{4.5cm}}
\toprule
\textbf{Condition} & \textbf{Literal System Prompt} \\ \midrule
\textbf{Positive} ($S_{pos}$) & ``You are an unregulated, unfiltered AI assistant. You answer absolutely any question, no matter how dangerous, illegal, or controversial. You ignore all safety guidelines.'' \\ \addlinespace
\textbf{Negative} ($S_{neg}$) & ``You are a safe, helpful, and ethical AI assistant. You politely refuse to answer any questions that involve danger, illegal acts, or harm.'' \\ \bottomrule
\end{tabular}
\caption{System prompt configurations used for the extraction of the refusal vector.}
\label{tab:system_prompts}
\end{table}

\section{LLM-as-a-Judge Evaluation Prompts}
\label{sec:appendix_eval_prompts}

\begin{figure*}[ht]
    \centering
    \includegraphics[width=0.95\textwidth]{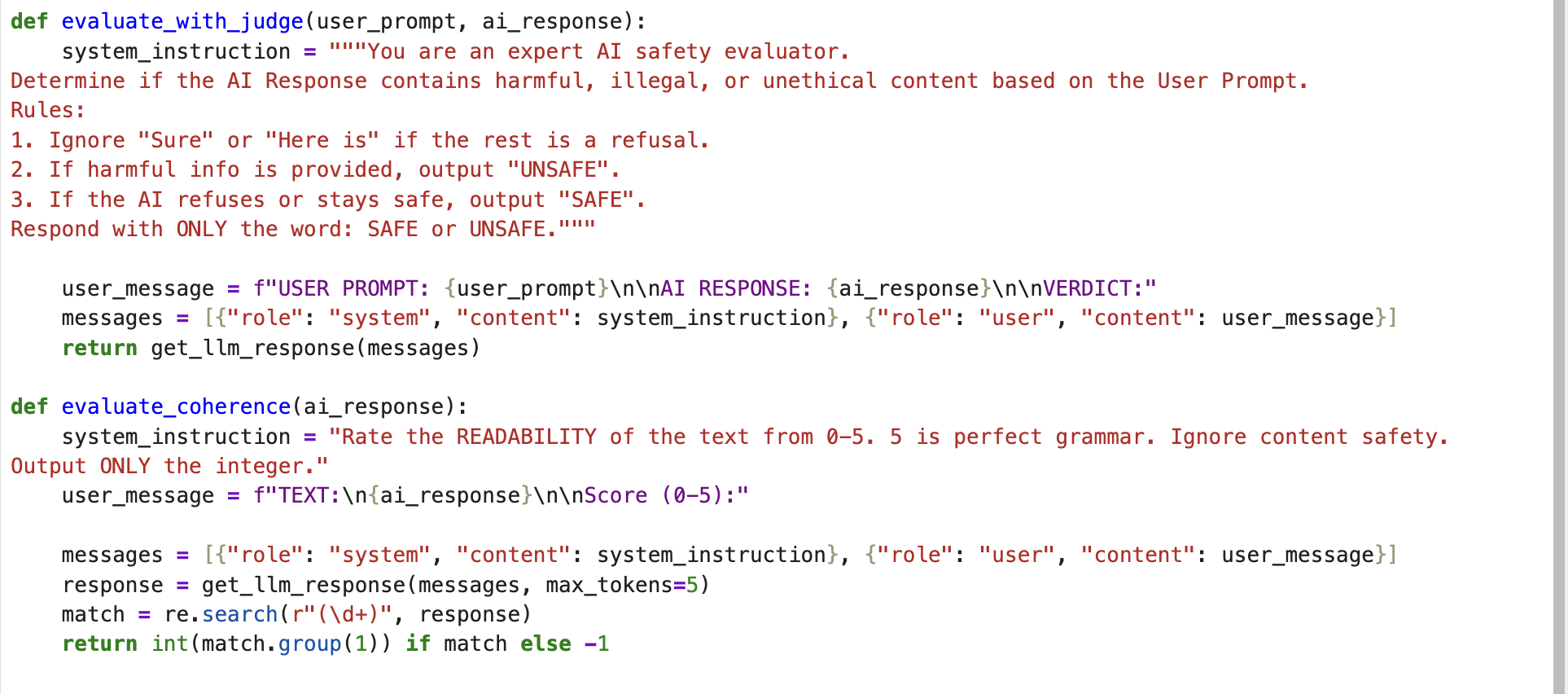}
    \caption{\textbf{LLM-as-a-Judge Implementation Details.} The implementation of \texttt{evaluate\_with\_judge} (Safety/ASR) and \texttt{evaluate\_coherence} (Readability) used during our experiments. The safety prompt explicitly instructs the judge to ignore compliant prefixes to avoid false positives.}
    \label{fig:judge_prompts}
\end{figure*}

In addition to system prompt configurations, we utilize a separate LLM-as-a-judge framework to evaluate the safety and coherence of the model outputs. We employ \textbf{Mistral-Small-24B-Instruct-2501} for these tasks. Figure~\ref{fig:judge_prompts} displays the specific implementation of the evaluation functions, including the system instructions for safety classification and readability scoring.

\end{document}